\documentclass{PoS}

\title{QCD at finite density}

\ShortTitle{QCD with $\mu\ne0$}

\newcommand\npart{N_{\rm part}}
\newcommand\lqcd{\Lambda_{\scriptscriptstyle QCD}}

\newcommand\bef{\begin{figure}}
\newcommand\eef{\end{figure}}
\newcommand\beq{\begin{equation}}
\newcommand\eeq{\end{equation}}
\newcommand\beqa{\begin{eqnarray}}
\newcommand\eeqa{\end{eqnarray}}
\newcommand\bet{\begin{table}}
\newcommand\eet{\end{table}}

\newcommand\fgn[1]{Figure \ref{fg:#1}}
\newcommand\eqn[1]{eq.\ \ref{#1}}
\newcommand\scn[1]{Section \ref{sc:#1}}

\newcommand\ie{{\sl i.e.\/}}

\newcommand\etal{{\sl et al.\/}}
\newcommand\jhep{{\sl J.\ H.\ E.\ P.\/}\ }
\newcommand\np{{\sl Nucl.\ Phys.\/}\ }
\newcommand\pl{{\sl Phys.\ Lett.\/}\ }
\newcommand\pros{{\sl PoS\/}\ }
\newcommand\pr{{\sl Phys.\ Rev.\/}\ }
\newcommand\prlt{{\sl Phys.\ Rev.\ Lett.\/}\ }

\author{\speaker{Sourendu Gupta}
            \\
        Dept.\ of Theoretical Physics, Tata Institute of Fundamental Research,
        Homi Bhabha Road, Mumbai 400005, India.\\
        E-mail: \email{sgupta@tifr.res.in}}

\abstract{Developments in QCD at finite density are reviewed. I begin by
discussing some new algorithms which have been applied to other theories
with sign problems.  Then I discuss the method of analytic continuation
in QCD using a series expansion and review some of the results obtained
using this method. By now there are several different simulations using
the method, and together they give estimates of the systematic lattice
effects, which turn out to be controlled.  Finally I discuss a direct
comparison of some of these lattice predictions with new experimental
data which results in a very pleasant agreement.}

\FullConference{The XXVIII International Symposium on Lattice Field Theory, Lattice2010\\
		June 14-19, 2010\\
		Villasimius, Italy}

\begin{document}

\goodbreak\section{Introduction}\label{sc:intro}

\bef
\begin{center}\includegraphics[scale=0.5]{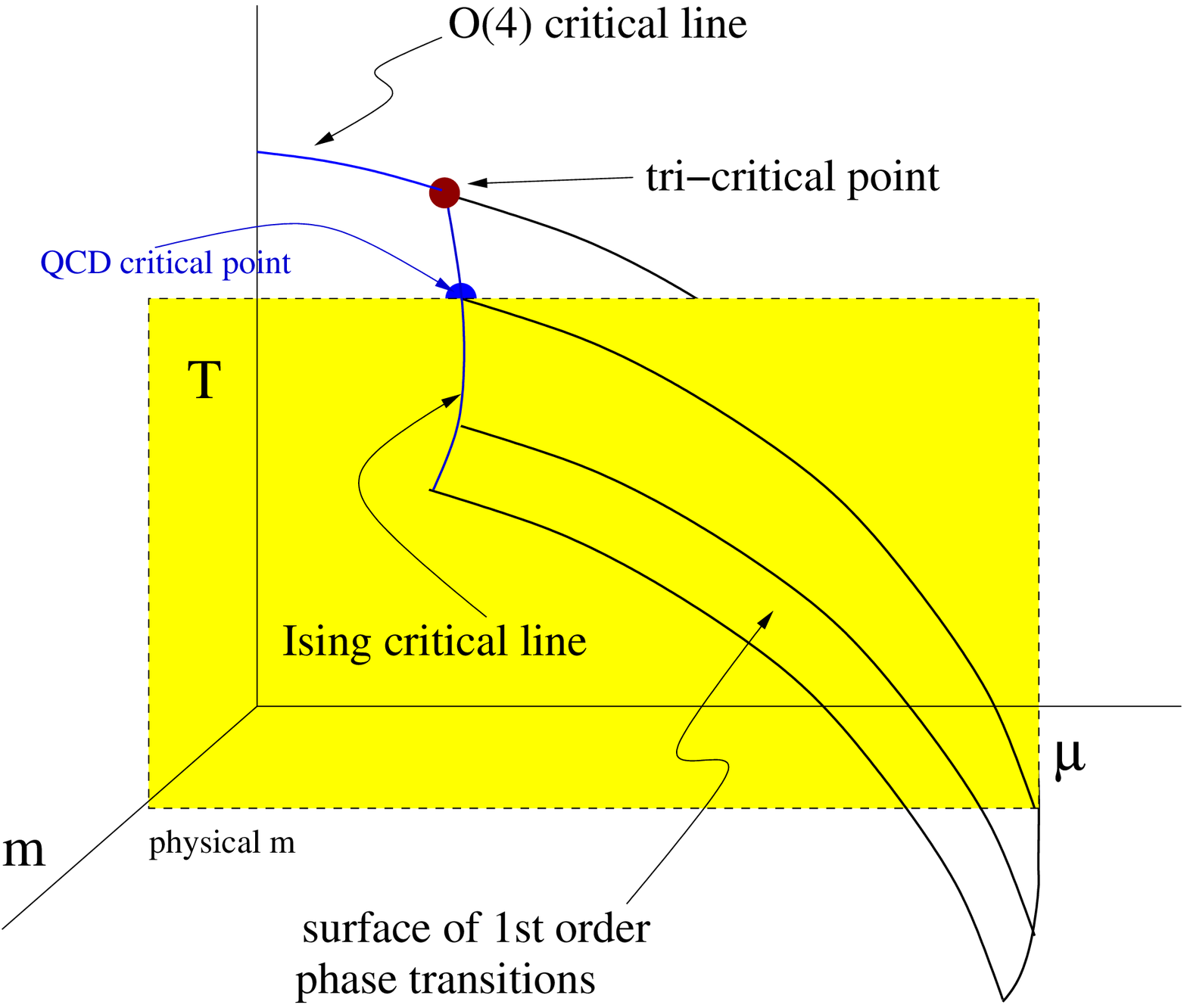}\end{center}
\caption{A conjectured phase diagram of QCD for $N_f=2$ and for $N_f=2+1$
when the strange quark mass is not much smaller than $\lqcd$. In the chiral
limit there is a tri-critical point, from which emerges an Ising critical
line whose intersection with the plane of physical quark mass is the QCD
critical point.}
\label{fg:qcdpd}\eef

QCD at finite baryon density is interesting because of two reasons: first
that there is a program of experimental studies covering five colliders,
running and planned, which will look at this problem, and
second, that it does not seem open to standard methods of attack in
lattice gauge theory due to a sign problem \cite{forcrand}.  In this
review I will bring together evidence that the problem is still open
to a fruitful attack using small modifications of the usual tools of
lattice gauge theory, and give some of the main physics results. The
context of these first results is the conjectured phase
diagram of \fgn{qcdpd} \cite{cep}.

Any Monte Carlo integration process suffers from a sign problem
if the integrand is not real and positive definite. For the QCD action with a
chemical potential on the baryon number, the determinant of the Dirac
operator, which is the quark part of the measure, obeys the condition
\beq
   \det(D+m+\mu\gamma_0)^* = \det(D+m-\mu^*\gamma_0),
\label{sign}\eeq
where $D$ is the massless Dirac operator, $m$ is the mass, $\mu$ is
the baryon chemical potential, and $*$ denotes complex conjugation. For any
generic complex chemical potential this shows that there is a sign problem.
For pure imaginary $\mu$ (including $\mu=0$), the determinant is real,
and one can further prove its positivity by considering its commutation
with $\gamma_5$.

This sign problem is not necessarily mild. Baryonless random matrix
theory seems to predict that for $\mu<m_\pi/2$ the distribution of signs is
Gaussian and becomes Lorentzian at larger $\mu$ \cite{ranmat}. In either
case the problem is severe.  An earlier work had estimated the contours
of the variance of the phase of the quark determinant and found that
this decreases at high temperatures, where the problem could therefore
become easier \cite{biswa2005}.

This review is structured as follows. \scn{algo} presents an overview
of very interesting new attempts to attack the sign problem directly;
unfortunately they are not yet at the stage where they can be applied to
QCD. \scn{series} reviews the Madhava-Maclaurin series expansion\footnote{
In the 14th century Madhava of Sangamagrama developed the series
expansion for functions and estimates of the error terms which later
came to be associated with the name of Maclaurin.} method which has
yielded first results on the phase diagram and on some other measurable
quantities. Finally in \scn{expt} first results from experiments are
reported along with comparisons to lattice QCD predictions.

\goodbreak\section{Trial algorithms}\label{sc:algo}

The class of algorithms which has had the most attention till now is
reweighting: perform the Monte Carlo procedure at a point in the phase
diagram where there is no sign problem, and then find expectation
values of operators by choosing an appropriate weight for each
configuration. Various problems with this process are by now well-known;
they include large errors due to cancellations and inaccurate sampling,
which become exponentially large with the volume. The Budapest group
applied this method to the problem of locating the QCD critical point
\cite{fk}. The Bielefeld-Swansea algorithm is a variant of this
method which expands the determinant in a series in $\mu$ \cite{biswa2002}.
There have been no developments in this class of algorithms since it
was reviewed in 2008 \cite{ejiri}.

\bef
\begin{center}\includegraphics[scale=1.0]{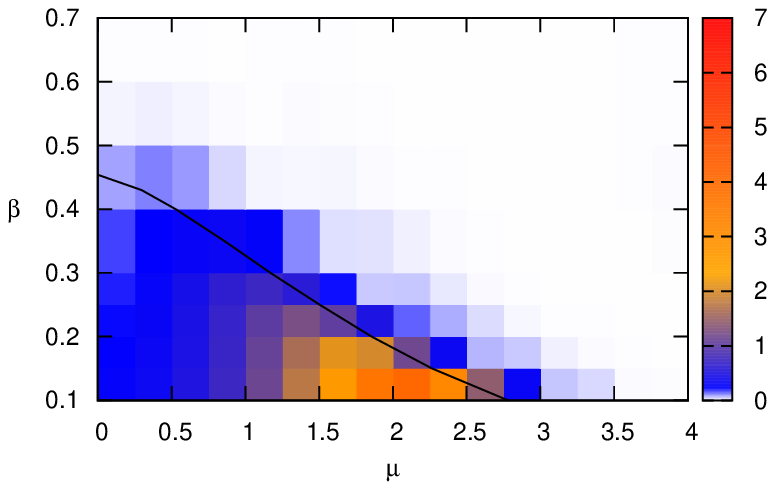}\end{center}
\caption{The line is the phase boundary of the 3d XY model, found using
the worm algorithm \cite{debasish}. The instability of the complex Langevin
method is illustrated in this figure \cite{aaja}: this method converges
to the wrong probability measure. The difference between the actions for the
true and converged distributions is colour coded.}
\label{fg:xy3d}\eef

Two new classes of algorithms are being tested currently, and, although
their applicability to QCD is not yet clear, they are interesting enough
to merit some discussion. Interestingly, they become easy to compare 
because the two algorithms have been, deliberately, applied to the same
model recently. This is the 3-d XY model at finite chemical potential,
which has the action
\beq
   S = -\beta \sum_{x,\hat\mu} \cos\left(\theta_x-\theta_{x+\hat\mu}
      -i\mu\delta_{\hat\mu,\hat t}\right).
\label{xy3d}\eeq
This suffers from a sign problem when $\mu\ne0$. 

One approach \cite{debasish} exploits the fact that sign problems are not
inherent to the physics of a system, but to specific representations. By
a clever transformation of fields which amounts to redefining the theory
in terms of fluxes of particles along links, they reduce it to a form
without a sign problem, although the theory then becomes non-local.
However, in this representation it becomes amenable to a numerical
attack using the ``worm algorithm'' \cite{wolff}. This work then sets
out a finite-size scaling theory which describes the point at which it
becomes energetically favourable to add one more particle to the ground
state. The simulation results allow the extraction of finite size scaling
 parameters which can then be used to determine the phase diagram.

The other approach resurrects an old idea--- the complex Langevin method,
wherein one addresses sign problems by complexifying the fields while
the noise remains real. Earlier works had been plagued by runaway
directions and associated numerical instabilities, now brought under
control by the use of adaptive step-size integrators. For a while
a proof of convergence of such methods seemed to be within reach
\cite{aaseist}. However, it turns out that there may be a convergence
to the wrong result \cite{aaja}. This is illustrated in \fgn{xy3d},
which shows that the problem arises mainly at small temperature and
large chemical potential. Since this region is similar to that in which
QCD has large sign fluctuations \cite{ranmat}, a better understanding
of the origin of this problem may throw light on applications to QCD.

In the next section we turn to the algorithm, first described in
\cite{ilgti2003}, which is now used by many groups, and has begun to
yield many consistency tests and, possibly, even contact with experiment.

\goodbreak\section{The Madhava-Maclaurin series expansion}\label{sc:series}

The pressure of QCD matter in a grand canonical ensemble can be expanded
in a Madhava-Maclaurin series around the point $\mu=0$ to obtain
\beq
   P(T,\mu) = P(T) + \frac{\mu^2}{2!}\chi^{(2)}(T) +
       \frac{\mu^4}{4!}\chi^{(4)}(T) + \cdots
\label{macl}\eeq
where all the coefficients are computed at $\mu=0$. $P(T)$ is the pressure
at zero chemical potential, $\chi^{(2)}(T)$ is called the quark number
susceptibility (QNS) \cite{milc1987} and all the $\chi^{(n)}(T)$ are
generically called non-linear susceptibilities (NLS). It was suggested
that the NLS could be measured in $\mu=0$ simulations, and the feasibility
was demonstrated by computations in quenched QCD \cite{ilgti2003}. More
recently, within the last year, there have been attempts to compute
these coefficients by simulating QCD at imaginary chemical potential
and fitting extrapolating functions to the data \cite{imagmu} (we will
return to a discussion of this later).

\subsection{Computational effort}

\bef
\begin{center}\includegraphics[scale=0.7]{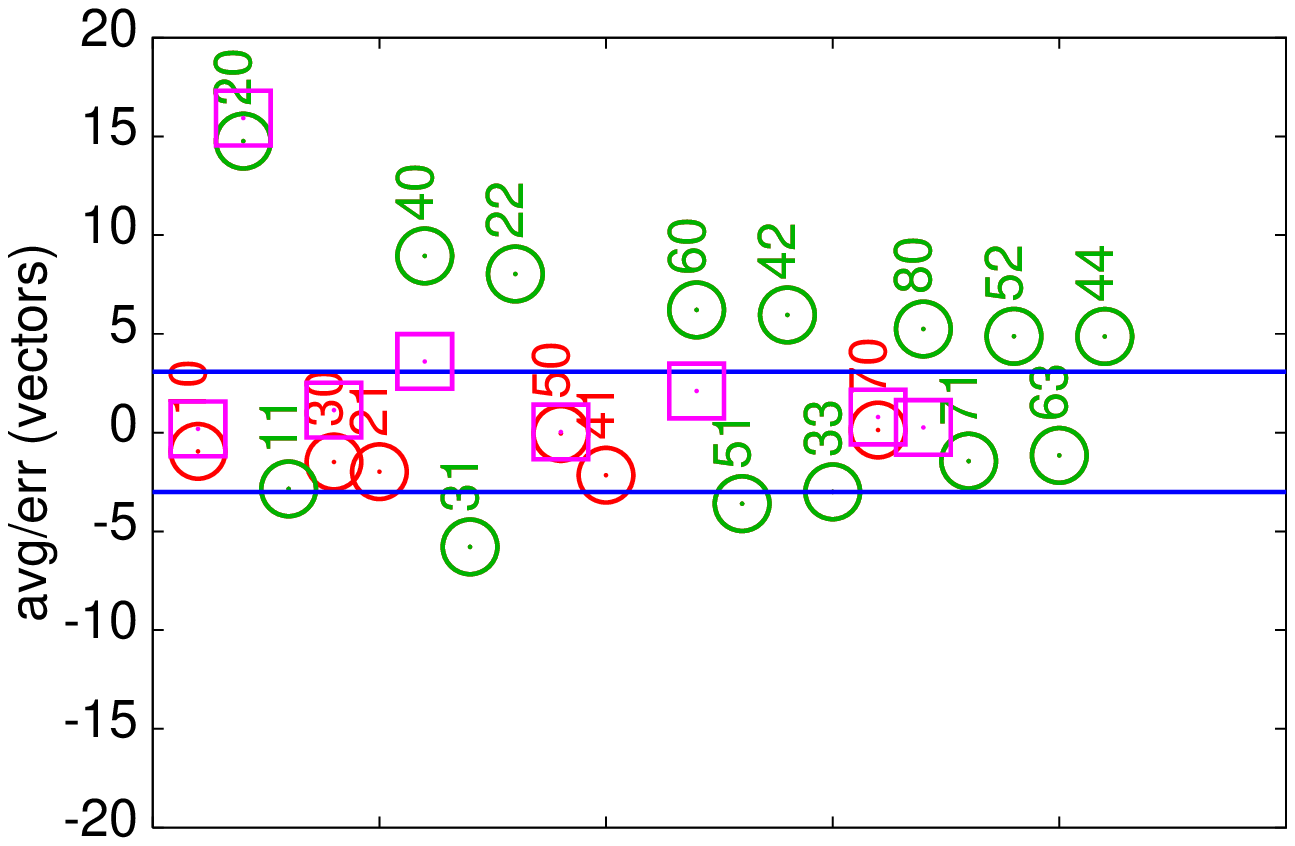}\end{center}
\caption{Signal to noise ratios for various fermion traces which enter
the evaluation of the NLS (see \cite{ilgti2005} for explanations of the
notation).  Circles denote data for staggered quarks \cite{ilgti2005} and
boxes for P4 quarks \cite{schmidt}. The red circles denote measurements
of quantities which should be exactly zero.}
\label{fg:signal}\eef

The $\chi^{(n)}$'s are combinations of quarks loops with insertions of
$\gamma_0$ up to $n$ times \cite{ilgti2005}. These quark loop traces
are obtained through stochastic noise averages. One measure of the
feasibility of such measurements is to examine the signal to noise ratio
in the measurements when the number of noise vectors is $N_v$, \ie, the
ratio of the mean and square root of the variance of such a trace in one
configuration. When the ratio is large, the measurement is easy. Such
a measure was reported using staggered quarks on a $4\times24^3$
lattice at $T/T_c=0.75$ and $N_f=2$ when the quark mass is tuned to give
$m_\pi=230$ MeV \cite{ilgti2005}. Here we add results from P4 action with
$N_f=2+1$, with light quark masses tuned to the same value of $m_\pi$
and $T/T_c=0.84$ \cite{schmidt}. In both cases $N_v=400$ and the signal
to noise ratios are comparable (see \fgn{signal}). A direct comparison
with Asqtad quarks is not available, but from the claim that 50\% of the
noise in \cite{milc2010} is due to stochastic estimators, one finds that
the signal to noise ratio for that Dirac operator is comparable.

Near $T_c$ autocorrelations between successive configurations is large---
of the order of 200--250 trajectories. Assuming that it takes about 200
fermion matrix inversions per trajectory, and that we use $N_v=500$ for
every decorrelated configuration, then, since it takes 18 inversions per
measurement (up to the 8th order of the expansion in eq.\ \ref{macl}), the
ratio of CPU times for a measurement to that for generating a decorrelated
configuration is 0.24. The marginal cost of measurement is small. Well
inside the hot phase, at $2T_c$, the autocorrelation time drops to about
4 trajectories, whereas $N_v=100$. The ratio of CPU times for measurement
to generating decorrelated configurations climbs to 4.5, however, with
relatively small expenditure of CPU time. As a result, direct measurements
of the NLS are highly feasible. An added attraction is that configurations
which have been generated for any finite temperature study can be reused
for such analysis, thus reducing the marginal cost even further.

\subsection{Series Analysis for the critical point}

Methods for analysis of series expansions of the free energy or its
derivatives are well-known in statistical mechanics, and have been used
successfully in many cases \cite{dandg}. While the core of the analysis
is the same, there are interesting differences between these older works
and the application to QCD, which lead to differences in the method of
analysis \cite{ilgti2005,ilgti2008}. The series coefficients in the older
works came from exact enumeration of graphs, corresponding to infinite
volume systems, so that a coefficient was either known exactly or not
known. In the present case the series coefficient is evaluated on finite
lattices with statistical errors through a Monte Carlo process.

The main point of the analysis is that the series for the quark number
susceptibility can be analyzed for its radius of convergence. The series
for the QNS is
\beq
   \frac{\chi^{(2)}(T,\mu)}{T^2} = \frac{\chi^{(2)}(T)}{T^2} +
       \frac{z^2}{2!}\chi^{(4)}(T) +
       \frac{z^4}{4!}T^2\chi^{(6)}(T) + \cdots
\label{chi2}\eeq
where $z=\mu/T$ and each of the dimensionless combinations
$T^{n-4}\chi^{(n)}(T)$ is the direct output of a lattice computation.
Since the expansion is in $z$ at fixed value of $T$, it is equivalent
to an expansion in $\mu$. If the divergence occurs at a temperature at
which all the series coefficients are positive, then the non-analyticity
occurs for real values of $z$, and the divergence can be identified with
the critical point of QCD.

\bef
\begin{center}
\includegraphics[scale=0.7]{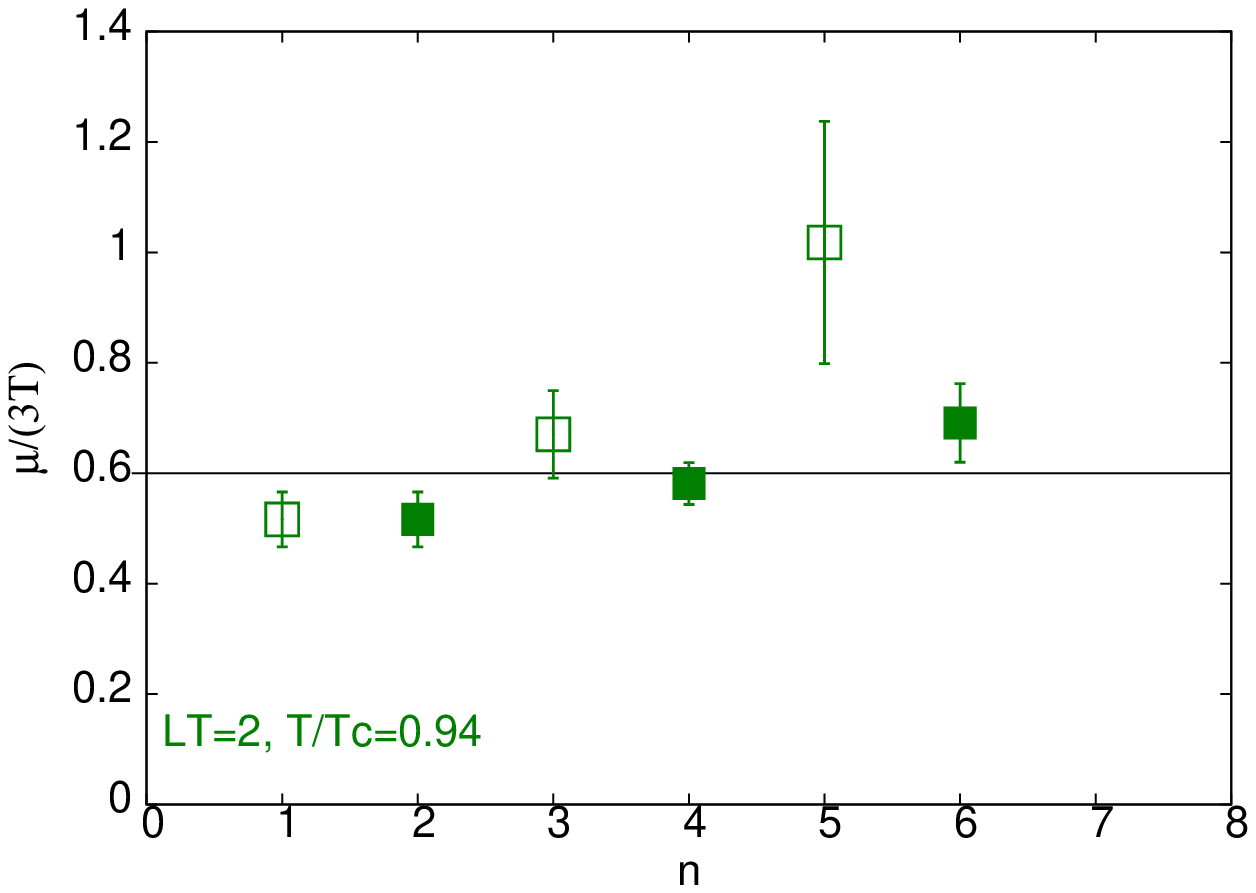}
\includegraphics[scale=0.7]{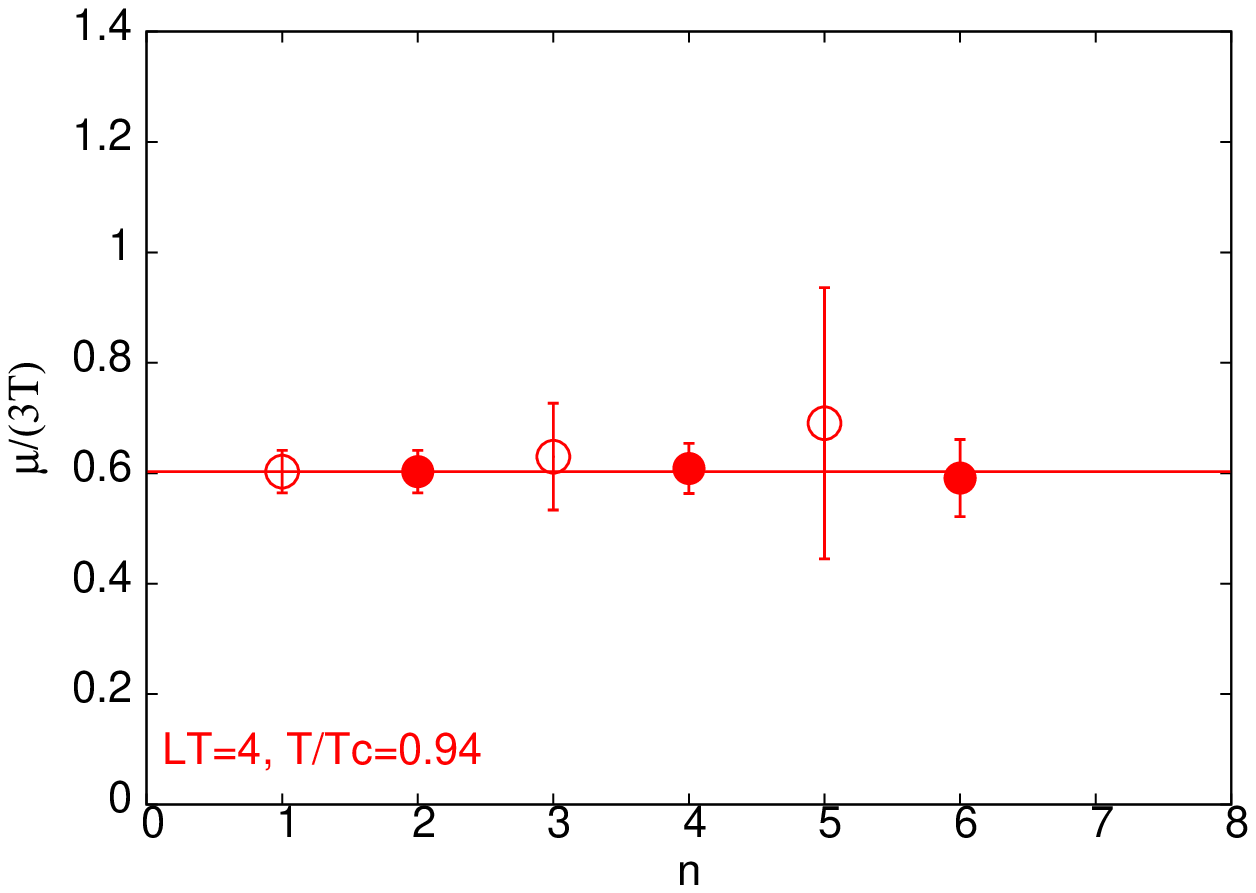}
\end{center}
\caption{Estimates of the radius of convergence of the series expansion in
\protect\eqn{chi2} from finite temperature simulations at $T=0.94T_c$ with two
flavours of staggered quarks and the bare quark mass tuned to give
$m_\pi=230$ MeV and lattice spacing $1/(6T)$. Open symbols correspond to
estimates of $z^*_n$ and filled symbols to ${\overline z}^*_n$. When $LT$
increases from 2 to 4 $n_*(L)$ increases significantly, as shown.}
\label{fg:fse}\eef

The radius of convergence can be found by several methods, all of which
correspond to comparing the series against another with a known singularity.
The best known definitions are---
\beq
   z^*_{n+1} = \sqrt{\frac{n!T^{n-2}\chi^{(n+2)}}{(n+2)!T^{n-4}\chi^{(n)}}},
   \qquad{\rm and}\qquad
   {\overline z}^*_n = \left(\frac{2!T^{n-4}\chi^{(n)}}{n!\chi^{(2)}/T^2}
          \right)^{1/(n-2)}.
\label{radc}\eeq
The star and bar do not indicate complex conjugation.
The common limit as $n\to\infty$ of both is the radius of convergence
of the series. This test is closely coupled to a finite volume
scaling analysis.

The reason is the following. If there is a critical point at some
$(\mu^E,T^E)$ then the QNS diverges there on an infinite volume
system. However, on any finite volume, $L^3$, there may be a peak, but no
divergence. As the system size decreases, the peak becomes broader and
lower. As a result, $z^*_n$ and ${\overline z}^*_n$ may seem to give a
finite radius of convergence for $n<n_*(L)$. For larger $n$ both $z^*_n$
and ${\overline z}^*_n$ will then become larger and larger, since there
is no actual divergence in the series for the QNS. As $L$ increases,
one would find $n_*(L)$ also increasing without limit.  In simulations of
QCD with 2 flavours of staggered quarks with the bare quark mass tuned to
give $m_\pi=230$ MeV, such behaviour is actually seen at one temperature
(see \fgn{fse}). At this temperature all the NLS are positive, so the
limiting singularity is at real values of $\mu$. We can then identify
such a temperature with $T^E$ and the corresponding estimate of the
radius of convergence with $\mu^E/T^E$.

Such estimates of the critical point have been made with two different
lattice spacings using two flavours of staggered quarks on large volume
lattices \cite{ilgti2005,ilgti2008}. A computation with 2+1 flavours of
P4 quarks at almost the same value of $m_\pi$ has also been performed
with large lattices \cite{rbrc2009} and preliminary estimates of the
radius of convergence have been reported \cite{schmidt2010}. These are
collected in \fgn{critpt}.  Since large volumes are crucial to obtaining
a stable estimate of the critical point, older computations with smaller
volumes have not been added into this figure even if they have realistic
values of $m_\pi$. A by-product of this choice is that all the points in
the figure use the same computational technique, albeit with different
lattice spacings and quark actions.

\bef
\begin{center}\includegraphics[scale=0.7]{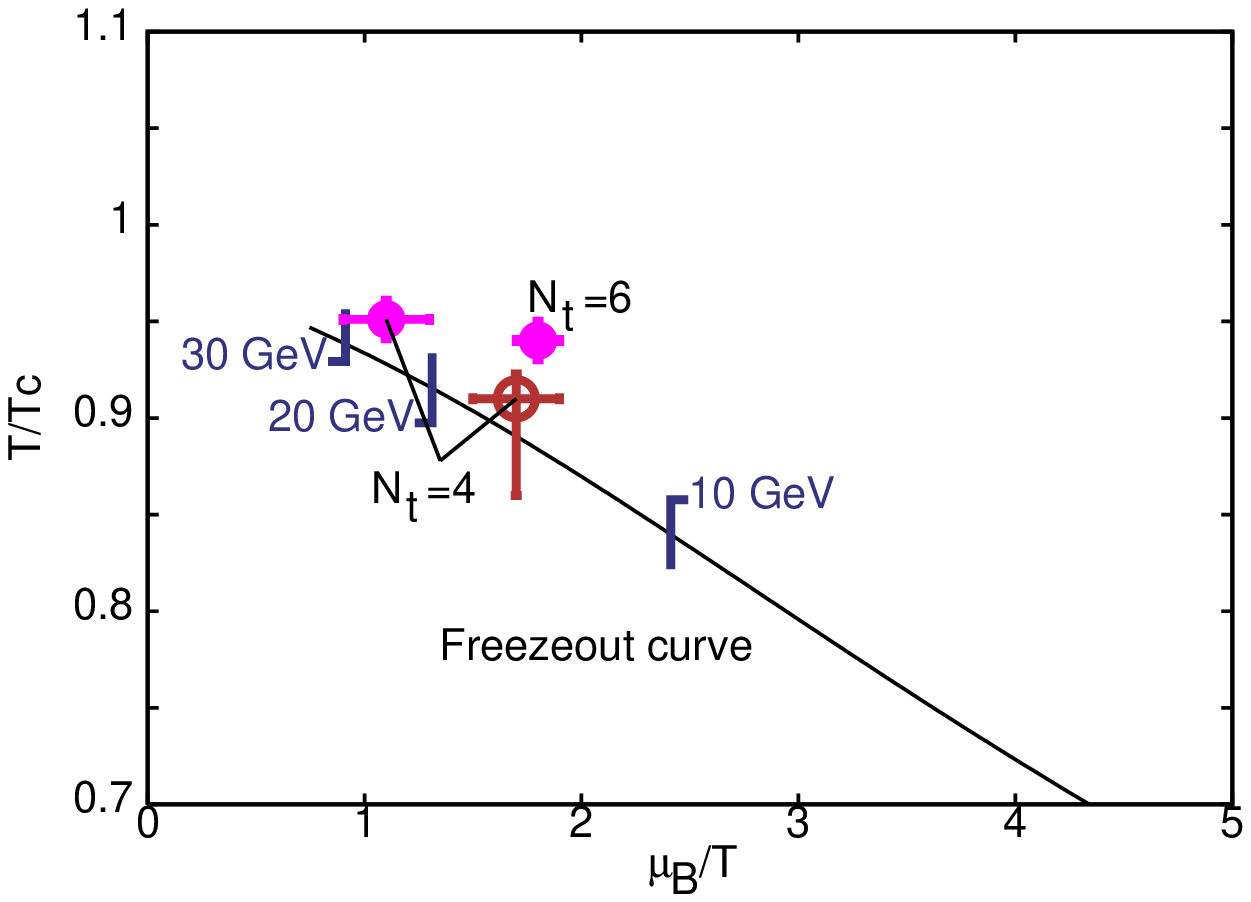}\end{center}
\caption{Estimates of the critical point of QCD from lattice computations
with $m_\pi\simeq230$ MeV and $Lm_\pi>4$. The points in pink are obtained
from computations with 2 flavours of staggered quarks and the point in brown
from 2+1 flavour P4 quarks. The values of $\mu$ and $T$ along the freezeout
curve are parametrized by \cite{cleymans} and have been used in conjunction
with $T_c=175$. Some values of $\sqrt{S}$ have been marked along the freezeout
curve.}
\label{fg:critpt}\eef

\bef
\begin{center}\includegraphics[scale=0.7]{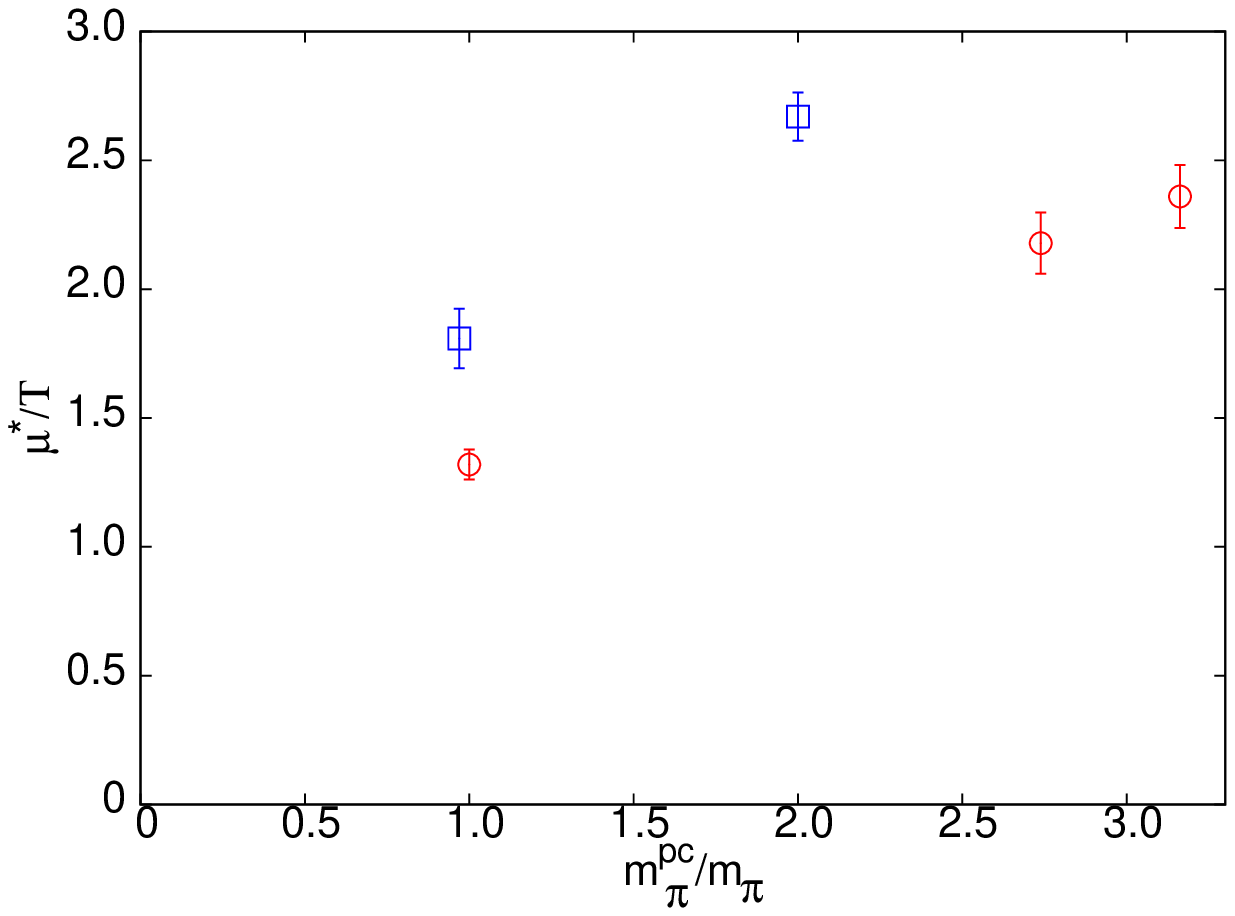}\end{center}
\caption{The variation of the radius of convergence with in partially
quenched computations with the staggered sea quark mass tuned to
give $m_\pi\simeq230$ MeV. As the valence quark mass is changed the
partially quenched pion mass is $m_\pi^{pc}$.}
\label{fg:massdep}\eef

Interest in the critical point is enhanced because of the possibility
that heavy-ion collision experiments may observe it. Fireballs produced
in such collisions undergo chemical freezeout at certain values of
$\mu$ and $T$ which change with the center-of-mass energy, $\sqrt{S}$,
of the colliding nuclei. The chemical freezeout point is relevant if one
tries to use fluctuations of conserved quantum numbers as probes of the
critical point: we shall return to this argument later. The freezeout
curve is parametrized in \cite{cleymans}. This has been superposed on
the phase diagram in \fgn{critpt} by using the scale $T_c=175$.

A pleasant fact emerges from \fgn{critpt}: that lattice spacing effects
can be bounded in magnitude by currently available computations. The
difference between different kinds of actions is, of course, a finite
lattice spacing artifact. The magnitude of the lattice spacing effect
estimated from two different spacings with the same action turns out to
be comparable with that from a comparison of two different actions at
nearly the same lattice spacing.

Interestingly, the effect of the strange quark on the end point seems
to be under control. It has long been known that in the Columbia plot,
the physical point corresponds to a thermal cross over \cite{columbia},
as a result of which the topology of the phase diagram of realistic 2+1
flavour QCD is the same as for two flavours, as in \fgn{qcdpd}. As the
strange quark mass is increased and the light quark mass is reduced,
the thermal crossover passes through a critical point into a first order
transition. It turns out that the line of critical points is far from
the physical point: the pion needs to be about 10 times lighter and the
strange quark mass about 3 times lighter \cite{strange}. This could be
one reason to suspect that the numerical values for the critical end
point in 2 flavour and 2+1 flavour QCD may not be very different. Such
an argument is compatible with the results collected in \fgn{critpt}.

The major remaining effect is due to the light quark mass being
larger than physical. The only exploration of this effect till now is
a partially quenched computations with $a=1/(4T)$ and two flavours
of staggered quarks with a sea quark mass tuned to give $m_\pi=230$
MeV \cite{sewm}. Interestingly, an interpolation between the measured
value of the radius of convergence was consistent with the result of
an unquenched computation with P4 quarks tuned to give $m_\pi=550$
MeV \cite{biswa2005}. In \fgn{massdep} the earlier results are extended
by adding a similar analysis for $a=1/(6T)$.  The extrapolation to the
physical value of $m_\pi$ shows a 15\% drop in the value of $\mu^E$.

\subsection{Extrapolation of observables}

\bef
\begin{center}\includegraphics[scale=0.35]{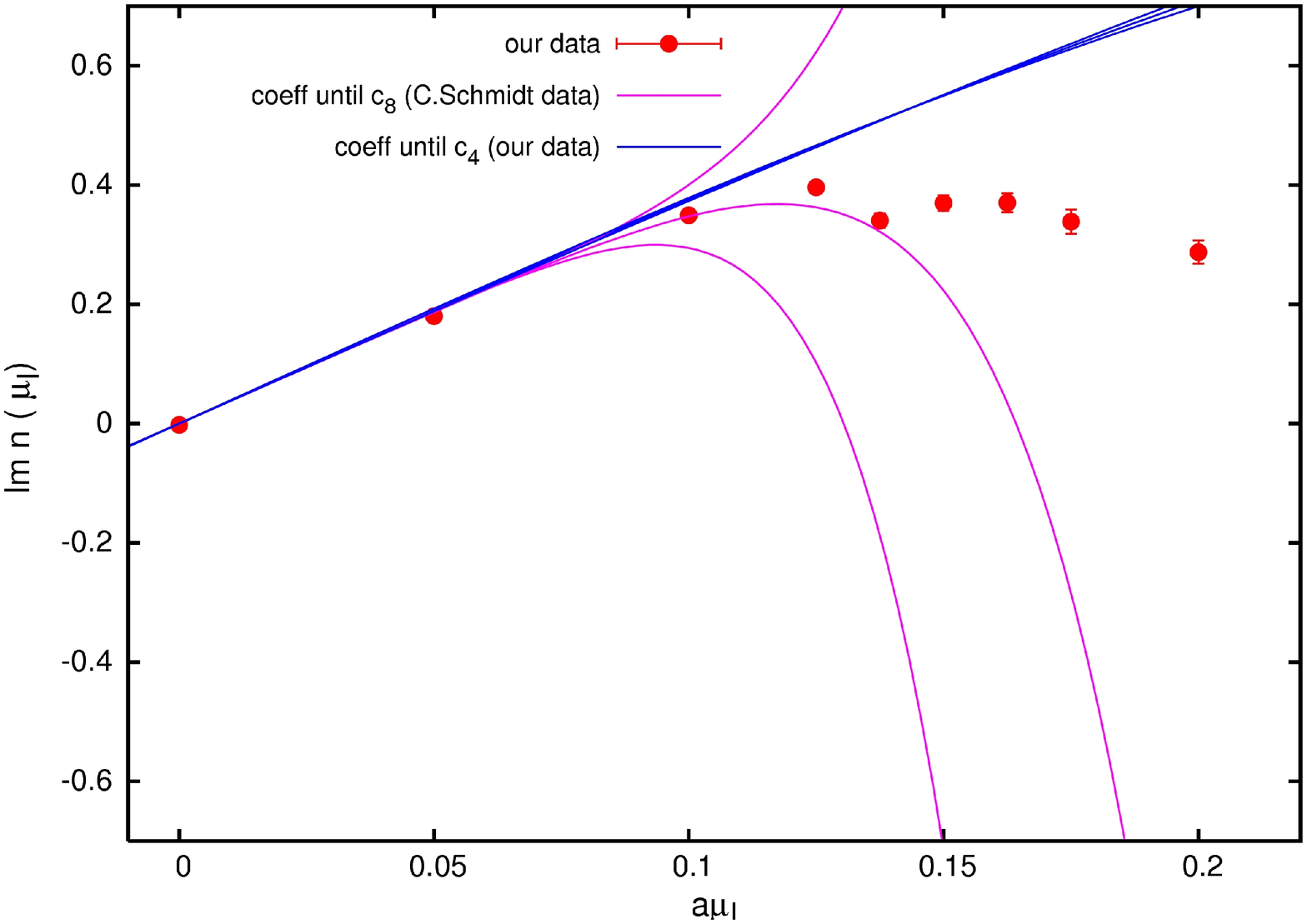}\end{center}
\caption{Using the series expansion to describe the data obtained through
direct simulations at imaginary chemical potential.}
\label{fg:falcone}\eef

Apart from the prediction of the critical point, the series
expansion could be used to extrapolate measurements to finite chemical
potential. Tests of such extrapolations are whether they can describe
measurements made directly through simulations at imaginary chemical
potential.  The most straightforward extrapolation is to use the series.
A preliminary attempt \cite{falcone} is shown in \fgn{falcone}. One sees
that adding new terms in the series improves the extrapolation only
marginally in $\mu$. Closely related to this exercise are attempts to
extract the series coefficients from measurements obtained in direct
simulations at imaginary chemical potential.  It was shown recently
\cite{imagmu} that simple series descriptions of the data obtained at
finite imaginary chemical potential are inefficient and more complicated
forms are needed to perform the extraction of the series coefficients.

\bef
\begin{center}
\includegraphics[scale=0.55]{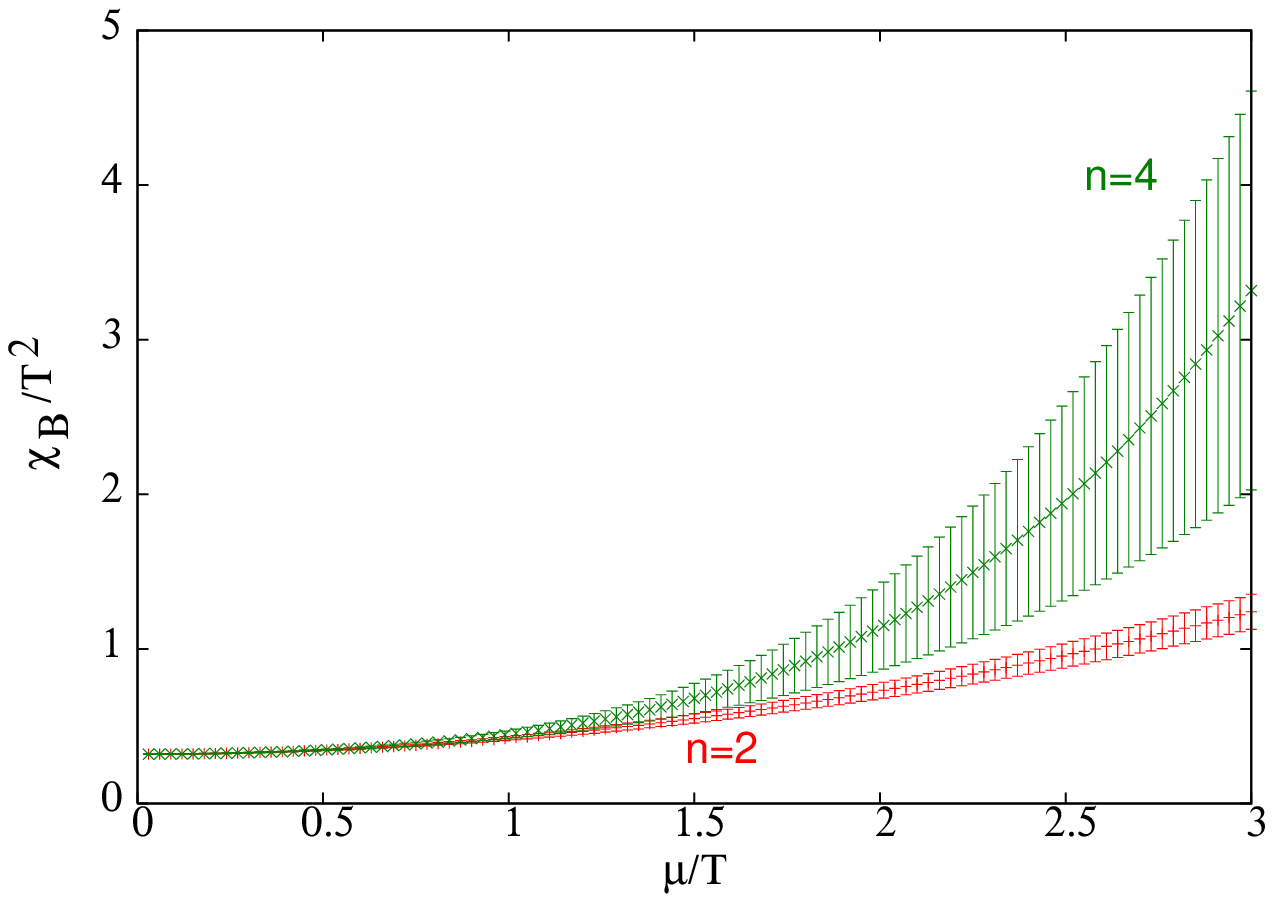}
\includegraphics[scale=0.55]{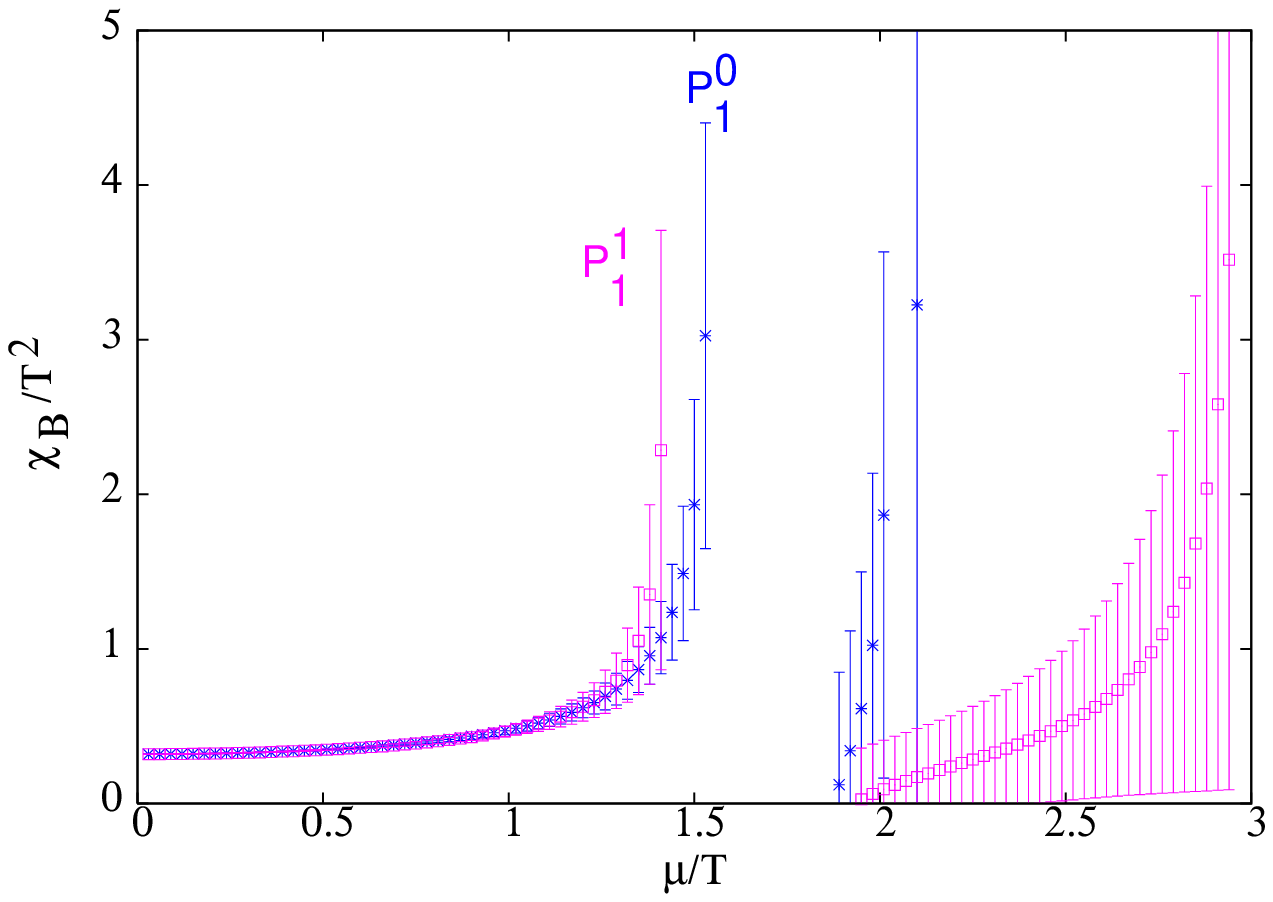}
\end{center}
\caption{Extrapolation of the measurement of the QNS at $T^E$ to finite
$\mu$. The series resummation does not diverge, although adding terms
causes large changes. The Pad\'e approximant exhibits the divergence but
is stable under addition of extra terms.}
\label{fg:pade}\eef

Clearly, when the series expansion of a physical quantity is close
to divergence, then a truncated sum is clearly not the best way to
find the value of this quantity at finite $\mu$. One must search for
series resummation techniques.  One method which has been widely used
for resummation of high temperature series of spin models \cite{dandg}
is to determine Pad\'e approximants using them.  There is a detailed
theory of Pad\'e approximants \cite{baker} which needs to be extended to
applications in QCD where the series coefficients are known only within
some statistical errors \cite{ilgti2008}.

In \fgn{pade} we show truncated series sums for the QNS at $T^E=0.94T_c$.
There is no sign of any divergence, although successive orders fail to
agree with each other as the radius of convergence is approached. In the
same figure we also show the QNS obtained with Pad\'e resummations of the
series. These exhibit the divergence identified through series analysis.
It is also useful to note that the Pad\'e approximants fitted to different
number of terms of the series agree with each other except when $z$
is significantly larger than the radius of convergence. The Pad\'e analysis
indicates a width of the critical region which is about $\Delta\mu/T^E\simeq
0.25$.

\goodbreak\section{Comparison with experiments}\label{sc:expt}

The QNS are related to fluctuations of conserved quantities in a grand
canonical ensemble. It may be possible to realize this in an experiment
by looking at a part of the fireball produced in a heavy-ion collision,
provided it thermalizes. Then one way in which grand canonical physics
can be extracted is by observing particles only in a restricted space-time
rapidity range. If this range is chosen judiciously, then the remainder of
the fireball may act as a heat-bath for the system under observation. Then
each collision event satisfying the above experimental cuts is one member
of a grand canonical ensemble of events.

Event-to-event distributions of conserved quantities then form the
observables of interest \cite{ebye}. The cleanest observable is the
distribution of total electric charge, $Q$, since there is very little
chance of missing a significant fraction of the charge within the
acceptance volume $V$. Baryon number, $B$, and strangeness, $S$, are
also good observables, but since there are uncharged baryons as well
as long-lived uncharged strange particles, both of which are missed by
detectors, the connection to these quantities is made at a further remove
\cite{stephanovhatta}.  Nevertheless, currently the most extensive data
comes from observations of the net proton number, which is a proxy for
the net baryon number.

It is seen that fluctuations of conserved quantities are Gaussian. The
first question is whether this Gaussian is entirely (or largely) due to
thermal fluctuations. The only way to answer this is by going beyond
the Gaussian. A systematic way to do this is to change $V$ and check
how the distribution changes.  Gaussian distributions usually arise in
experiments through a process described by the central limit theorem: with
increasing $V$ the higher cumulants of the distribution scale down with
larger powers of $V$. The STAR collaboration reported such a measurement
\cite{starqm2009} using an experimentally determined parameter called
the participant number, $\npart$, as a proxy for $V$. At small $\npart$
all the cumulants, $[B^n]$, are comparable, and with increasing $\npart$
the scaling of the cumulants is exactly as one should expect--- in other
words, the microscopic physics encoded in the set of $[B^n]$ does not
change with $\npart$.

\bef
\begin{center}\includegraphics[scale=0.35]{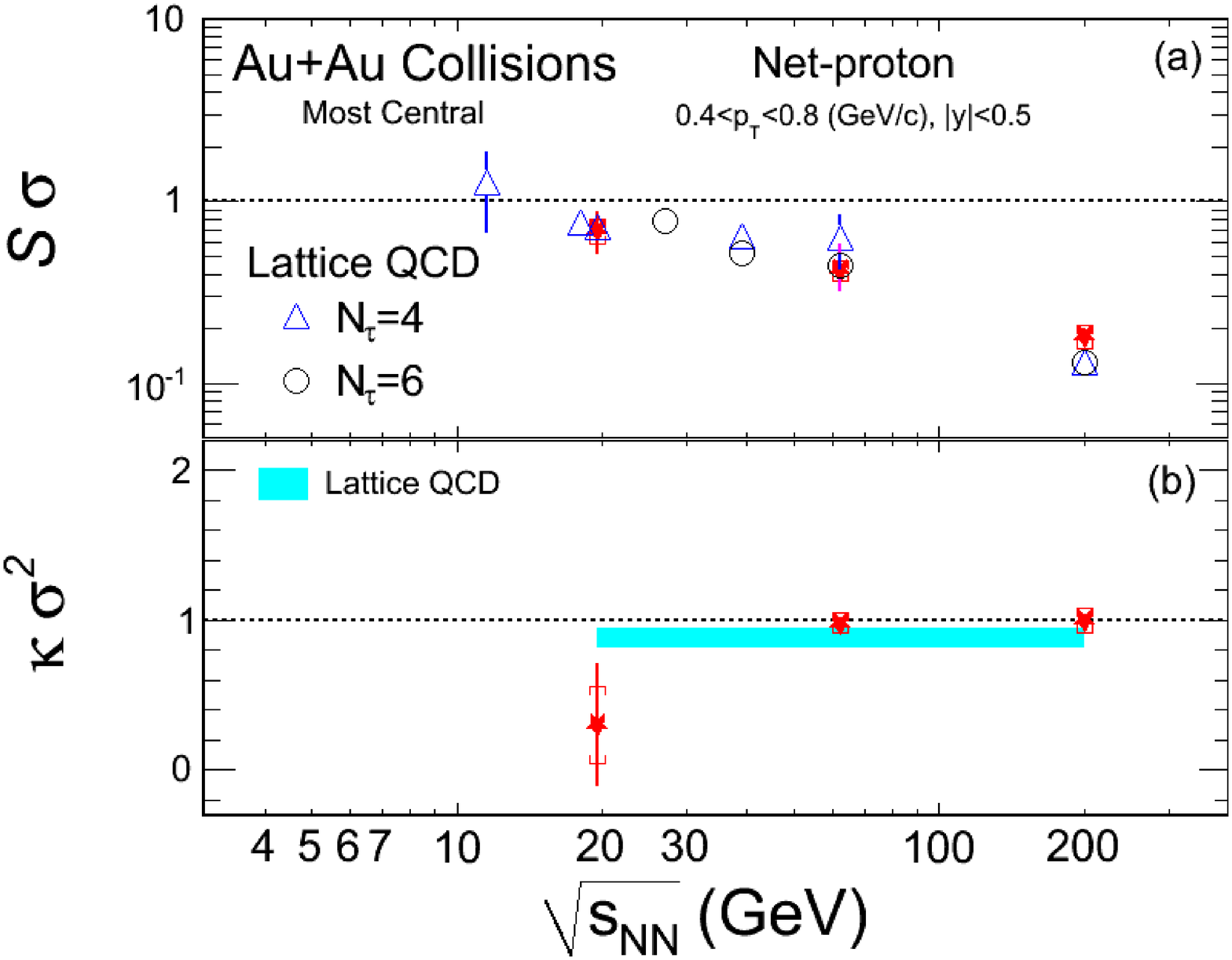}\end{center}
\caption{Comparison of experimental data with lattice predictions. The
upper panel shows $m_1$ and the lower $m_2$.}
\label{fg:star}\eef

If the fluctuations are due to thermal processes, then $[B^n]$
are related to various NLS computable in QCD. This is the
next step: to check the data against the predictions \cite{cpod2009}. In
order to do this one has to take combinations of the cumulants which
are free of incidental variables such as the unmeasurable $V$.
This is achieved by taking the ratios
\beq
   m_1=\frac{[B^3]}{[B^2]}=\frac{T\chi^{(3)}(T,\mu)}{\chi^{(2)}(T,\mu)},\quad
   m_2=\frac{[B^4]}{[B^2]}=\frac{T^2\chi^{(4)}(T,\mu)}{\chi^{(2)}(T,\mu)},\quad
   m_3=\frac{[B^4]}{[B^3]}=\frac{T\chi^{(4)}(T,\mu)}{\chi^{(3)}(T,\mu)},
\label{meas}\eeq
Now, the left hand side of each equation is known from experiment at each
$\sqrt S$, and the right hand side is known from lattice computations
if one knows the freezeout values of $\mu$ and $T$. These values are
parametrized from experimental data in \cite{cleymans} assuming that the
fireball thermalizes.

On one hand, the series expansions for the NLS are known from the
expansion in \eqn{macl}, on the other, the ratios of \eqn{meas}
have well-determined power behaviour at small $z$, and poles near
the critical point. As a result, resummation of the series by Pad\'e
approximants is possible. The parameters in the Pad\'e approximants are
closely related to the estimates of $z^*_n$.  As a result, the lattice
artifacts in $m_{1,2,3}$ are related to those already discussed in
the previous section.  It turns out that both $m_1$ and $m_2$ may have
significant finite lattice spacing corrections: mainly a common finite
multiplicative factor which is also the correction to the estimate to
the radius of convergence. For $m_3$ lattice spacing corrections are
small, except in the vicinity of the critical point.  Results for these
quantities have been given in \cite{ilgti2010}.

The STAR experiment at RHIC has recently published a measurement of
comparable quantities from runs at three different values of $\sqrt S$
where comparisons with these quantities are presented \cite{star2010}.
As shown in \fgn{star}, it turns out that there is good agreement between
the data and the predictions. Many questions remain to be answered: on the
side of the lattice computations the usual questions about flavours, quark
masses and lattice spacing, on the experimental side about the removal of
non-thermal backgrounds and other sources of fluctuations.  Nevertheless,
this is a significant milestone: the first direct comparison of heavy-ion
data with lattice predictions.  In future such a comparison may even yield
a direct measurement of $T_c$ as pointed out in \cite{ilgti2010}, allowing
us to set the scale for lattice measurements in an entirely new way.

\goodbreak\section{Assorted topics}

There are various developments at finite chemical potential which
cannot be covered fully here because of space constraints. However,
they are interesting in their own right and have useful connections to
the physics which is discussed in the previous sections. Here I make a
brief mention of some of these works.

In the chiral limit there is a line of second order phase transitions
at finite $\mu$, emanating from the finite temperature critical point
(see \fgn{qcdpd}). The curvature of this line is an object of interest
because it sets a scale for the tricritical temperature in the chiral
limit. Interesting new work was presented on this problem by several
groups \cite{curv}.

The phase diagram at imaginary chemical potential is of some interest,
since it has to be understood if simulations in this region are to
be used for analytic continuation to the physically interesting case.
New results were presented by two groups \cite{newimag}.

The investigation of correlation functions at finite chemical potential
is in its infancy \cite{oldcorr}. Any new information is interesting at
this stage.  New work was reported in this meeting \cite{newcorr}.

The strong coupling expansion has been resurrected in this context
and improved techniques are now being used to investigate the phase
diagram. Interesting new results in this direction were reported
\cite{strcpl}.  These may serve to benchmark future simulations using
the worm algorithm which can be adapted to the strong coupling theory.

All the results reported till now simulate the grand canonical
ensemble. Very little systematic effort has gone into simulations in
the canonical ensemble with fixed baryon number. One such attempt was
reported \cite{canonical}.

\goodbreak\section{Conclusions}

Over the years there has been a great improvement in the understanding of
the sign problem at finite $\mu$ in QCD: where it could be tractable and
where it is not \cite{ranmat}. There has been little progress in directly
tackling this problem, although there are interesting developments in the
handling of other models with sign problems \cite{debasish,aaja}. However,
in the last five years there has been enormous progress in lattice
computations which can yield information on QCD at finite $\mu$. The
essential development is the use of analytic continuation, through a
Madhava-Maclaurin series expansion in $\mu$ \cite{ilgti2003}. There are
encouraging tests of this method in performing analytic continuation
to finite imaginary $\mu$ where direct simulations are also possible
\cite{imagmu,falcone}.

The series expansion method has been applied to the
extraction of the QCD critical point at various lattice spacings
\cite{ilgti2005,ilgti2008} and with various quark actions and numbers of
flavours \cite{rbrc2009,milc2010}. A composite figure of the predictions
is given in \fgn{critpt}.  It is clear from this figure that the method
yields results with controlled statistical and systematic errors. Since
the results are not strongly sensitive to the choice of action, it is
also clear that lattice spacing effects are bounded.

An interesting development in the past year has been the proposal of a set
of measurables, \eqn{meas}, which allow a direct comparison of experiment
and lattice computations. First results show very good
agreement between data \cite{star2010} and prediction \cite{ilgti2010}.
This calls for renewed activity in this field and a greater scrutiny of
the known systematic uncertainties which need control.

For communicating their results and then patiently answering my questions
I would like to thank Gert Aarts, Shailesh Chandrasekharan, Rossella
Falcone, Maria-Paola Lombardo, and Christian Schmidt.

\end{document}